\documentclass[aps,prl,twocolumn,showpacs]{revtex4}
\usepackage{bm}
\usepackage{graphicx}
\usepackage{amsmath}
\usepackage{eufrak}
\usepackage{color}
\newcommand{\nix}[1]{}
\begin{document}

\title{Spin polarized electric currents in semiconductor heterostructures\\ induced by microwave radiation}
\author{C. Drexler,$^1$ V.V. Bel'kov,$^2$ B. Ashkinadze,$^3$ P. Olbrich,$^1$ C. Zoth,$^1$
V. Lechner,$^1$
Ya.V. Terent'ev,$^2$ D.R. Yakovlev,$^{2,4}$
G. Karczewski$^5$, T. Wojtowicz,$^5$ D. Schuh,$^1$  W. Wegscheider,$^1$  and S.D.\,Ganichev$^{1}$
}
\affiliation{$^1$  Terahertz Center, University of Regensburg, 93040
Regensburg, Germany}
\affiliation{$^2$A.F.\,Ioffe Physical-Technical Institute, Russian
Academy of Sciences, 194021 St.\,Petersburg, Russia}
\affiliation{$^3$ Solid State Institute, Technion-Israel Institute
of Technology, 32000 Haifa, Israel}
\affiliation{$^4$ Experimental Physics 2, TU Dortmund University, 44221 Dortmund, Germany,}
\affiliation{$^5$ Institute of  Physics, Polish Academy of Sciences, 02668 Warsaw, Poland}

\begin{abstract}
We report on microwave (mw) radiation induced electric currents in
(Cd,Mn)Te/(Cd,Mg)Te and InAs/(In,Ga)As quantum
wells subjected to an external in-plane magnetic field. The
current generation is attributed to the spin-dependent energy
relaxation of electrons heated by mw radiation. The relaxation
produces equal and oppositely directed electron flows in the
spin-up and spin-down subbands yielding a pure spin current. The
Zeeman splitting of the subbands in the magnetic field leads to
the conversion of the spin flow into a spin-polarized electric
current.
\end{abstract}

\maketitle

The discovery of  microwave induced oscillations in the
resistivity of a two-dimensional electron gas (2DEG) attracted
growing attention to an electron magneto-transport in
semiconductor nanostructures subjected to 
mw radiation,
see, e.g.,\,\cite{Zud,Kuku}.
The
experimental observation of mw-induced effects stimulated much
theoretical interest\,(see \cite{Dmit} and references therein)
since they provide information which is complementary to conventional transport.
In addition, the mw-induced effects offer new ways for
developing sensitive microwave detectors
\,\cite{Kukushkin2}. All these effects have been observed
applying an external magnetic field  perpendicularly to a
2DEG plane.

In this Letter we demonstrate that microwave radiation induces a
novel electron transport effect also under an in-plane magnetic
field, i.e., in the geometry that excludes the cyclotron motion
and Landau quantization of the two-dimensional electrons. The
effect is caused by an asymmetric spin-dependent electron energy
relaxation of the 2DEG  heated by mw
radiation\,\cite{Naturephysics,PRL09}. In an external magnetic
field, this process results in a spin-polarized electric current.
The mw-induced currents are observed in two different
semiconductor systems: diluted magnetic semiconductor (DMS)
(Cd,Mn)Te/(Cd,Mg)Te quantum wells (QWs) and  InAs/(In,Ga)As QWs.
In these structures the spin-polarized electric currents are
enhanced due to either the presence of magnetic Mn$^{2+}$
ions\,\cite{PRL09} or the strong spin-orbit coupling in InAs
\,\cite{Naturephysics}.

\begin{figure}[ht]
\includegraphics[width=0.75\linewidth]{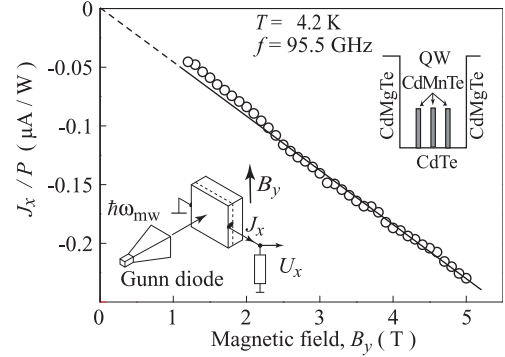}
\caption{Magnetic field dependence of the mw-induced signal $J_x$
for DMS (Cd,Mn)Te/(Cd,Mg)Te QW sample (circles). Insets show the
experimental geometry and the structure sketch. } \label{B_depend}
\end{figure}

Two   \textit{n}-type doped QW structures have been grown by
molecular-beam epitaxy on (001)-oriented 
substrates. The
first sample has a digital alloy DMS QW\;\cite{Jaroszynski2002}, which is
a 10\,nm-wide 
CdTe QW containing three monolayers of Cd$_{0.86}$Mn$_{0.14}$Te 
(see  inset in Fig.\,\ref{B_depend}). The effective average concentration of
Mn 
is $\bar{x}=0.013$. The sample is
$\delta$-doped with Iodine donors introduced into the top
Cd$_{0.76} $Mg$_{0.24} $Te barrier at a 15\,nm distance from the
QW. The 2DEG has the density $n$= 6.2$\times
$10$^{11}$\,cm$^{-2}$ and the mobility $\mu =1.6\times
$10$^4$\,cm$^2$/Vs at a 
temperature $T = 4.2$\,K.
For further properties 
see
Ref.\,\cite{PRL09}. The second sample is a non-magnetic 
InAs QW consisting of a 20\,nm wide 
In$_{0.75}$Ga$_{0.25}$As/In$_{0.75}$Al$_{0.25}$As structure with a 4~nm InAs channel
embedded in InGaAs.
It is Silicon
$\delta$-doped with $n =1\times $10$^{12}$\,cm$^{-2}$ and  $\mu
 =7\times $10$^4$\,cm$^2$/Vs in the 2DEG at $T=4.2$\,K.

\begin{figure}[ht]
\includegraphics[width=0.78\linewidth]{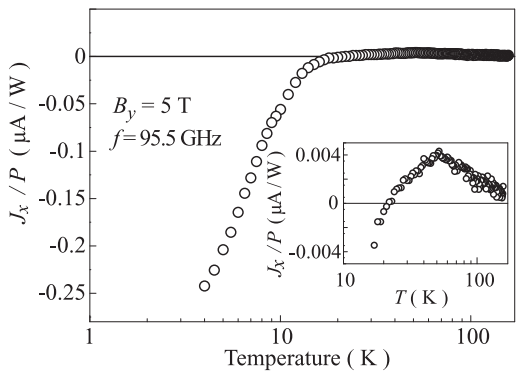}
\caption{Temperature dependence of the mw-induced photocurrent
$J_{x}$  in the DMS (Cd,Mn)Te/(Cd,Mg)Te QW. Inset shows the zoom
of the zero crossing region. } \label{T_depend}
\end{figure}

In the mw-induced current experiments,  samples of $5\times
5$\,mm$^2$ size with a pair of ohmic contacts along the direction
$x \parallel [1\bar {1}0]$ are used. The experimental geometry is
sketched in the inset of Fig.\,\ref{B_depend}. The samples were
placed into an optical cryostat and a split-coil  superconducting
magnet yielding a field $B_{y} \parallel [110]$ with a strength up
to 5\,T. The sample temperature was varied from 1.8\,K to 150\,K.
The microwave radiation was generated by two Gunn oscillators
operating at frequencies, $f$, of 95.5\,GHz or 60\,GHz. The
radiation is guided through the cryostat windows on the sample at
normal incidence utilizing a horn antenna and two parabolic
mirrors. The distance between the horn antenna output and the
sample was about 320\,mm. The incident mw power of about 0.1\,mW
was modulated by a p-i-n switch at 970\,Hz. The mw-induced current
in the unbiased structure, $J_x$ (perpendicular to $B_{y}$) was
measured with lock-in technique via the voltage drop $U_x$ across
a 1\,M$\Omega$  or 50\,$\Omega$ load resistors.

Figure\,\ref{B_depend} shows the current $J_{x}$ induced
by 95.5\,GHz irradiation of the DMS 
QW  
at $T=4.2$\,K  
as a function of the magnetic field strength.
The current increases 
with $B_{y}$
and reverses its sign as the direction of  $B_{y}$ changes (not
shown). As the lattice temperature increases from 2\,K to 50\,K,
the current  value decreases by more than two orders of magnitude
and the signal reverses its sign at $T\approx 20$\,K, see
Fig.\,\ref{T_depend}. Similar results have  also been obtained for
the mw-frequency of 60\,GHz.

\begin{figure}[ht]
\includegraphics[width=0.75\linewidth]{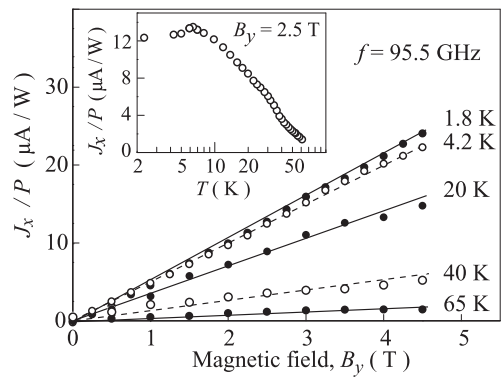}
\caption{Magnetic field dependence of the mw-induced electric
current $J_x$ in the InAs/(In,Ga)As QW. Its temperature dependence
is shown in the inset. } \label{InAs}
\end{figure}

Figure\,\ref{InAs} demonstrates  $J_{x}(B_{y})$ dependences for
the non-magnetic InAs/(In,Ga)As QW obtained at 95.5\,GHz
irradiation for several $T$. The mw-induced current is
detected  in the temperature range from $1.8$\,K
 to 65\,K.
The current  depends linearly on $B_{y}$,
however, contrary to the DMS sample, it decreases smoothly with rising
$T$ and does not change the sign (see  inset in
Fig.\,\ref{InAs}). 
It will be shown below that such a significant difference in the
$J_x (T)$ dependences is caused by various temperature behavior of
the Zeeman spin splitting in both materials, which for DMS
structures is strongly influenced by the polarization of Mn
ions\,\cite{spinbook,Dietl}.

\begin{figure}[t]
\includegraphics[width=0.6\linewidth]{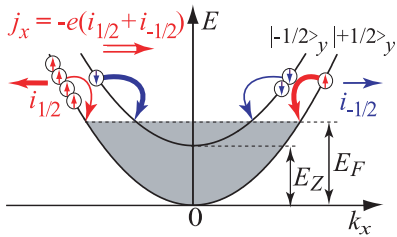}
\caption{Model of the mw-induced generation of spin-polarized
current.
See details in the text.} \label{Figure5model}
\end{figure}

Our findings can be well understood in the frame of the recently
proposed model for the spin-dependent asymmetric energy relaxation
of a nonequilibrium (mw-heated) electron
gas\,\cite{Naturephysics}. Free electrons absorbing microwave
radiation, are excited to higher energy states,  then they relax
into an equilibrium state by emitting phonons.
Figure\,\ref{Figure5model} sketches the  energy relaxation
processes of hot electrons in two spin subbands ($s_y =\pm 1/2)$
which are split in an external magnetic field due to the Zeeman
effect. The  energy relaxation
is shown by bend arrows. Usually, the energy relaxation via
scattering by phonons is considered to be spin-independent.
However, in gyrotropic media, like CdTe- and InAs-based QWs, the
spin-orbit interaction adds an asymmetric spin-dependent term to
the scattering probability with the matrix element being
proportional to $[\bm{\sigma}\times(\bm{k}+\bm{k}^\prime)]$ (note,
that we assume presence of the structure inversion asymmetry
only)\,\cite{Naturephysics}. Here, $\bm{\sigma}$ is the axial
vector composed of the Pauli matrices; $\bm{k}$ and
$\bm{k}^\prime$ are the initial and final electron wave vectors.
Thus, the electron transitions to positive and negative
$k_x^\prime $-states occur with different probabilities leading to
an imbalance in the distribution of carriers between positive and
negative $k_x$-states  in both subbands ($s_y =\pm 1/2)$. This is
shown by bent arrows of different thicknesses for both subbands in
Fig.\,\ref{Figure5model}. In the absence of an external magnetic
field the asymmetry of the electron-phonon interaction results in
a spin current, but not in an electric current since the
oppositely directed electron fluxes, $i_{\pm 1/2} $, in spin
subbands $s_y =\pm 1/2$ are of equal strength and, therefore,
compensate each other. The external magnetic field leads to the
Zeeman splitting of the $s_y$ subbands. As a result, the
electron densities in the spin-up and spin-down subbands 
become different, and the fluxes $i_{\pm 1/2} $ do not longer
compensate each other yielding a net electric current. Obviously,
the current is spin polarized and its value is proportional to the
Zeeman splitting energy, $E_{Z}$. Such a physical mechanism of the
mw-induced current generation explains the significantly different
behavior of the currents in the DMS and non-magnetic structures.

In DMS structures the exchange interaction of electrons with
magnetic ions results in the giant Zeeman spin splitting  of the
conduction band, $E_{Z}$, and amplifies  spin-dependent phenomena,
see, e.g.,\,\cite{spinbook,Dietl,Fur88}. Moreover, $E_{Z}$ has a
remarkable temperature dependence \cite{Fur88}

\begin{equation}
\label{eq2} E_{\rm Z} = g \mu _{\rm B} B +  \bar{x} S_0 N_0
\alpha
 {\rm B}_{5/2} \left( {\frac{5 \mu _{\rm B} g_{Mn} B}{2 k_{\rm B}
(T_{Mn} + T_0 )}} \right) \, ,
\end{equation}

where $k_{\rm B} $ is the Boltzmann constant and  $\mu _{\rm B}$
the Bohr mag\-neton. The first term is the intrinsic spin
splitting with the electron $g$-factor $g\,=\,-1.64$ for CdTe
QWs\,\cite{Sir97}. The second one describes the electron exchange
interaction with Mn$^{2+}$ ions, where $g_{Mn}\,=\,2$ is the Mn
$g$-factor and $T_{Mn}$ the Mn-spin system temperature. The
parameters $S_0$ and $T_0$ account for the Mn-Mn antiferromagnetic
interaction, $\rm{B}_{5/2}$ is the modified Brillouin function and
$N_0 \alpha\,=\,220$\,meV is the
exchange integral
\,\cite{Dietl}.
The sign inversion of  $J_x (T)\propto E_Z$,  shown in the inset of Fig.\,\ref{T_depend}, directly follows from
Eq.\,(\ref{eq2}). 
Indeed,  $E_{Z}$, reverses its sign upon temperature
variation due to opposite signs of $g$ and $N_0 \alpha$. This
explains the experimental data shown in
Fig.\,\ref{T_depend}. It is worthwhile to note that in our previous photoluminescence
study, the sign inversion of the Zeeman splitting was detected at
$T\,\approx\,40$\,K\,\cite{PRL09}.
This temperature is higher than
the inversion temperature detected in the mw-induced current
 ($T\,=\,20$\,K).
We attribute  this difference to the mw-heating of the Mn spin
system above $T$, as reported in\,\cite{PRL09,Kel02}.

In contrast to our DMS QWs, the sign of the Zeeman splitting in the
non-magnetic InAs-based structure ($E_{\rm Z} = g \mu _{\rm B}
B$) does not change (for bulk InAs $g \approx
-15$\,\cite{InAsgfactor}). Hence, the 
temperature dependence of
the mw-induced current shown in the inset of Fig.\,\ref{InAs} is
mostly determined by the spin up/down subband population
(Boltzmann factor)\,\cite{Naturephysics}. We notice, that a
similar $J_x (T)$ behavior was also observed 
at terahertz
(THz) frequencies\,\cite{Naturephysics,PRL09}. The mechanism
of the current generation for THz- and mw-excitation is the same,
however, in the mw-range the electron gas heating for a fixed power
is higher  due to the larger Drude-absorption for smaller photon energies.

The signals detected in our experiments, in particular at low
temperatures, are rather large (being of the order of $\mu$V) and
therefore, may contribute substantially to the magneto-transport
phenomena studied under in-plane magnetic fields\,\cite{Du}. For
(001)-grown zinc-blende QWs, the spin-polarized   current
generation can only occur under application of an
\textit{in-plane} magnetic field: The photocurrent generation with
$\bm{B}$ normal to the 2DEG plane geometry ($\bm{B}  \parallel z$)
is forbidden by symmetry reasons. However, in [110]- and
[113]-grown QWs, the effect becomes allowed for the $\bm{B}
\parallel z$-geometry\,\cite{PRL08}
 and, therefore, may be of importance for the analysis of
mw-induced current-voltage effects studied in magnetic fields
applied perpendicularly to 2DEG plane.  
To summarize, our work demonstrates a convenient method
of spin polarized currents generation 
by microwave radiation of conventional Gunn diodes.
The direction and magnitude of such
spin-polarized currents can be controllably varied by a proper
combination of the crystallographic orientation and magnetic field
direction as follows from the discussed physical mechanism for the
current generation.

\acknowledgments  The financial support from the DFG (SFB 689 and SPP 1285),
the Linkage Grant of IB of BMBF at DLR,
Russian Ministry of Education and Sciences, and RFBR, is gratefully acknowledged.
The research was partially supported by the EU within European
Regional Development Found, through grant Innovate Economy
(POIG.01.01.02-00-008/08) and by the Foundation for Polish Science
through subsidy 12/2007.

\end{document}